\documentstyle{mn}
\input psfig.sty
\newcommand{\ltsima} {$\; \buildrel < \over \sim \;$}
\newcommand{\gtsima} {$\; \buildrel > \over \sim \;$}
\newcommand{\lta} {\lower.5ex\hbox{\ltsima}}
\newcommand{\gta} {\lower.5ex\hbox{\gtsima}}

\def\b{\beta}

\def\d{\delta}

\def\refitem{\par\parskip 0pt\noindent\hangindent 20pt}

\title[RELATIVISTIC LARGE SCALE JETS AND MINIMUM POWER REQUIREMENTS] 
{Relativistic large scale jets and minimum power requirements}

\author[G. Ghisellini, A. Celotti]
{G. Ghisellini$^1$ and A. Celotti$^2$\\
$^1$ Osservatorio Astronomico di Brera, V. Bianchi, 46, I-23807 Merate,
Italy\\
$^2$ S.I.S.S.A., V. Beirut 2--4, I-34014 Trieste, Italy}

\begin{document}
\maketitle

\begin{abstract}
The recent discovery, by the $Chandra$ satellite, that jets of blazars
are strong X--ray emitters at large scales (0.1--1 Mpc) lends support
to the hypothesis that emitting plasma is still moving at highly 
relativistic speeds on these scales.
In this case in fact the emission via inverse Compton scattering off 
cosmic background photons is enhanced and the resulting predicted X--ray 
spectrum well accounts for the otherwise puzzling observations.  
Here we point out another reason to favour relativistic large scale jets, 
based on a minimum power argument: by estimating the Poynting flux and 
bulk kinetic powers corresponding to, at least, the relativistic particles 
and magnetic field responsible for the emission, one can derive the value 
of the bulk Lorentz factor for which the total power is minimized. 
It is found that both the inner and extended parts of the jet of
PKS~0637--752 satisfy such a condition.
\end{abstract}


\section{Introduction}

Most of the radiation observed from blazars (BL Lac objects and flat
spectrum radio quasars) originates in relativistic jets, in which the
emitting plasma flows with a bulk Lorentz factor $\Gamma \sim 10$ at
small viewing angles $\theta$.  Since the observed radiation is beamed
in the direction of motion the large apparent luminosities are much
reduced in the comoving frame, by a factor $\Gamma^4$ (if $\theta\sim
1/\Gamma$) in the case of a moving blob.  On the other hand, the most
important power content of the jet is not in the form of internal
random particle energy giving rise to radiation, but in the form of
electromagnetic and kinetic powers associated with the bulk motion.
This is indicated also by the power budget of the extended radio
structures and radio lobes, which require an average supply which is
greater than the luminosity generated by the jet at all scales.
However, the jet does not need to remain relativistic for all its
length, and indeed it was believed that it slows down, without losing
much of its bulk kinetic energy content.  
In fact the detection of large--scale, symmetric jets has been 
considered as an indication that on such scales the plasma is moving 
at most with mildly relativistic velocities in low power FR~1 sources 
(Fanaroff \& Riley 1974), while the jet to counterjet flux ratio requires
$\beta c \sim$0.6--0.8 in FR~II large--scale jets (e.g. Laing 1993; 
Bridle et al. 1994; Wardle \& Aaron 1997; Hardcastle et al. 1999). 
However, as these authors (and Laing et al. 1999) pointed out, there is 
the possibility of a 
velocity structure across the jet, whose plasma could move highly
relativistically close to the axis (``spine") and mildly
relativistic at the border (``slow layer"). 
As found by Chiaberge et al. (2000), Celotti et al. (2001), such a 
velocity structure can also help to unify radio--galaxies seen at
large viewing angles and blazars.

Observations of large scale jets emitting copious amounts of X--rays also
suggest that jets might maintain ultra--relativistic speed up to the
largest scales.  
We consider as a specific case that of the source PKS
0637--752, 
observed by {\it Chandra},
for which observational information is available on both large and
small scales. {\it Chandra} observations reveal relatively intense 
X--ray emission quite far from the nucleus (Chartas
et al. 2000).  The spatially resolved X--ray images show knots of
emission similar to those seen in radio maps (Tingay et al. 1998) and
optical images (HST; Schwartz et al. 2000).  VSOP observations
detected superluminal motion in the jet associated with the source
(Lovell 2000), setting a lower limit on the (inner, i.e. pc scale)
bulk Lorentz factor $\Gamma >17.5$ and an upper limit on the viewing
angle $\theta< 6.4$ degrees (assuming $H_0=50$ km $s^{-1}$ Mpc$^{-1}$
and $q_0=0$).  Furthermore the small scale jet appears to be well
aligned with the kpc scale one: by de--projecting the X--ray map using
an angle of 6 degrees the observed X--ray emission extends
up to $\sim$1 Mpc from the nucleus.

As discussed by Chartas et al. and Schwartz et al., on large scales
the radio spectrum can be explained as synchrotron emission, as also
supported by the detection of strong linear radio polarization, while
the X--ray emission exceeds by a large factor ($\sim$ 2 orders of magnitude) 
the extrapolation from the radio--optical spectrum.  If the X--ray flux is
synchrotron self--Compton (SSC) radiation from plasma moving at most at
mildly relativistic speed, then large deviations from equipartition
between magnetic field and particle energy density and/or large
inhomogeneities have to be invoked in the emitting region.  On the
other hand thermal emission by e.g. shocked plasma also appears to be
unlikely, as it would imply high particle density and thus unobserved
large rotation measures.  
On the other hand the X--ray flux can be
satisfactorily interpreted as scattered Cosmic Microwave Background
radiation (CMB) by relativistic plasma moving with a bulk Lorentz
factor $\Gamma\sim$10--15 (Ghisellini 2000; Celotti, Ghisellini \&
Chiaberge 2001; Tavecchio et al. 2000).  
These observations then suggest that the jet, or some part of it, can 
indeed maintain highly (bulk) relativistic speeds from milli-- to 
megaparsec.

Here we introduce another argument in support of large bulk Lorentz
factors along the entire jet extension.  This is based on minimizing
the power required to produce the radiation we observe, once we
consider not only the dissipated radiative luminosity, but also the
Poynting flux and the particle bulk kinetic power of the jet.  Note
that, as the power in bulk motion dominates over the internal particle
energy, the minimum power criterion does not coincide with the
equipartition argument usually applied to radio lobes (Burbidge 1959).
We show that the behaviourof the total power as function of the bulk
Lorentz factor depends on the dominant emission process (chosen to
reproduce the intrinsic quantities) and has a minimum for a certain value
of $\Gamma$.

In the following we first explicitly derive the condition for the
minimum total power requirement (Section 2).  This is applied to the
X--ray extended emitting knot and small scale jet of PKS 0637--752 in
Section 3.  The implications for the energy transport in jets and 
some observational predictions are discussed in Section 4.

\section{The minimum total power}

The spectral energy distribution (SED) of blazars and more generally
the emission by non--thermal plasma in relativistic jets can
usually be represented by one or two broad components. 
These are commonly interpreted as synchrotron (the low energy one) 
and possibly inverse Compton emission (the high energy one, if present) 
from a population of relativistic leptons. 
The seed photon population for the inverse--Compton scattering may be the
synchrotron photons themselves (synchrotron self--Compton, SSC) or 
photons produced externally to the emitting plasma.

Let us assume the simplest scenario and approximate the emitting
region as a spherical `blob' of plasma of size $R$, in bulk motion
with Lorentz factor $\Gamma$ at an angle $\theta$ with respect to the
line of sight. This produces an intrinsic power $L^\prime_{\rm s}$ in
synchrotron radiation, and $L^\prime_{\rm c}$ in inverse Compton
radiation. These are the luminosities as measured in the comoving
frame.  The corresponding dissipated luminosities (i.e. the power
received by a detector completely surrounding the source and which
sees it moving with $\Gamma$) are enhanced by the factor
$\sim(4/3)\Gamma^2$.  Instead the `observed' luminosities $L_{\rm s}$
and $L_{\rm c}$, estimated by multiplying the received flux by $4\pi
d_L^2$ ($d_L$ is the luminosity distance), are a factor $\delta^4$
greater then the intrinsic ones. Here
$\delta=[\Gamma(1-\beta\cos\theta)]^{-1}$ is the Doppler factor and
$\beta c$ the blob bulk velocity.

The plasma comprises leptons (protons) of comoving density $n_{\rm e}$
($n_{\rm p}$), embedded in a magnetic field of strength $B$
perpendicular to the direction of motion.  
The power carried as Poynting flux and the kinetic powers corresponding to 
the bulk motion of leptons and protons are respectively given by:
\begin{eqnarray}
L_{\rm B}\, &=& \, {1 \over 8} R^2 \Gamma^2 \beta c B^2\, ,\nonumber \\  
L_{\rm e}\, &=& \, \pi R^2 \Gamma^2 \b c  n_{\rm e} 
      \langle \gamma \rangle m_{\rm e} c^2 \, ,\nonumber \\  
L_{\rm p}\, &=& \, \pi R^2 \Gamma^2 \b c  n_{\rm p} m_{\rm p} c^2.
\end{eqnarray}
Here $\langle \gamma \rangle$ is the average random Lorentz factor of
the relativistic leptons, measured in the comoving frame, and $m_{\rm
p}$, $m_{\rm e}$ are the proton and electron rest masses,
respectively.  Protons are assumed to be cold in the comoving frame.

We can estimate these quantities from the observed radiation.  
A lower limit on $n_{\rm e}$ can be inferred from the synchrotron
emission $L_{\rm s}$ of the plasma. In fact the number density of
leptons producing the observed radiation is:
\begin{equation}
n_{\rm e} \, = \, {9 \over 2 \sigma_{\rm T} c } {L_{\rm s}
   \over \langle \gamma^2 \rangle \d^4 R^3 B^2}, 
\end{equation}
where $\langle \gamma^2 \rangle$ is averaged over the relativistic
electron distribution and $\sigma_{\rm T}$ is the Thomson scattering
cross section.  
The particle bulk kinetic powers can then be expressed as
\begin{eqnarray}
L_{\rm e}\, &=& \, 
     {9\pi \langle\gamma\rangle m_{\rm e} c^2 \over 
     2 \sigma_{\rm T} \langle \gamma^2\rangle } \,
     {\Gamma^2 \beta \over \delta^4 }\, {L_{\rm s} \over R B^2},  \nonumber \\     
L_{\rm p}\, &=& \, {9\pi m_{\rm p} c^2 \over 
     2 \sigma_{\rm T} \langle \gamma^2\rangle } \,
     {n_{\rm p}\over n_{\rm e}}\,
     {\Gamma^2 \beta \over \delta^4 }\, {L_{\rm s} \over R B^2}.
\end{eqnarray}
This implies that for a given size of the emitting region, $\Gamma$,
$\theta$, observed synchrotron luminosity and spectrum (which
determines $\langle\gamma\rangle$ and $\langle\gamma^2\rangle$), the
total power $L_{\rm B}+L_{\rm e}+L_{\rm p}$ must have a minimum for
some value of $B$.

A similar approach has been adopted for the case of the galactic
superluminal source GRS~1915$+$105, for which $\Gamma$ and $\theta$
are known, by Gliozzi, Bodo \& Ghisellini (1999).  In the case of
blazars the viewing angle is constrained by selection effects to be
close to $\theta=1/\Gamma$, since the flux is strongly dimmed for
larger viewing angles, while the solid angle
becomes too small for $\theta < 1/\Gamma$. 
Under this further
condition, the total power depends only on the product $\Gamma B$.  
In fact in this case $\delta \sim \Gamma$, and then $L_{\rm B}$ and
$L_{\rm e}+ L_{\rm p}$ behave in opposite ways with respect to $\Gamma
B$. The value corresponding to the minimum total power can be easily
determined by setting ${\partial/\partial (B\Gamma)} (L_{\rm e}
+L_{\rm p} + L_{\rm B})\equiv 0$, to derive $(B\Gamma)_{\rm min}$:
\begin{equation}
(B\Gamma)_{\rm min}\, =\, \left[{36\pi L_{\rm s}m_{\rm e} c^2\over 
  \sigma_{\rm T} c \langle\gamma^2\rangle R^3}\,
  \left(\langle\gamma\rangle  + 
 {n_{\rm p} m_{\rm p} \over n_{\rm e} m_{\rm e}} \right)\right]^{1/4};\,\,\,\, 
 \delta=\Gamma.
\end{equation}
The behaviours of the different powers and the resulting minimum of the
total one are illustrated, as an example, in Fig.~1.  Here we
represent their values as a function of $\Gamma B$ in the case of
``normal" electron--proton ($e$--$p$) plasma and pure
electron--positron ($e^\pm$) pair plasma.
\begin{figure}
\psfig{file=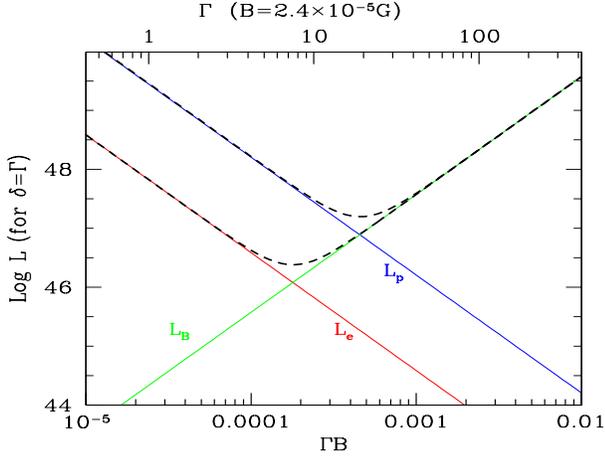,width=9truecm,height=7truecm}
\vskip -0.5 true cm
\caption{
The kinetic (in protons, $L_{\rm p}$, and in leptons, $L_{\rm e}$)
and magnetic powers as functions of $\Gamma B$,
the product of the the bulk Lorentz factor and the magnetic field. 
The viewing angle is assumed such that $\delta=\Gamma$. The axis on the
top refers to values of $\Gamma$ for a representative value of B.}
\end{figure}

Interestingly a second observational constraint allows us to decouple
$B$ and $\Gamma$. In fact the ratio of the luminosity in the inverse
Compton component to that in the synchrotron component (hereafter the Compton
dominance) is proportional to the ratio of the relevant energy
densities, in radiation and magnetic field.

Consider first the case of inverse Compton on photons produced
externally to the jet.  The ratio between the Compton and synchrotron
luminosities can be expressed as
\begin{equation}
{L_{\rm c} \over L_{\rm s} } \, 
= \,  8\pi U_{\rm ext} {\Gamma^2 \over B^2}
\end{equation}
where the radiation energy density 
$U^\prime_{\rm ext}$ ($\sim \Gamma^2 U_{\rm ext}$
for viewing angles close to $1/\Gamma$ as in our case; 
see Dermer 1995), 
in the comoving frame of 
a source at redshift $z$ includes e.g. the CMB radiation field for large 
scale structures $U^\prime_{\rm CMB}= aT^4(1+z)^4 \Gamma^2$ ($T=2.7$K is the
temperature we observe now), and/or the photons of the Broad Line
Region on sub-pc scales $U^\prime_{\rm BLR}$, etc. Thus the magnetic
field required to give the observed $L_{\rm c}/L_{\rm s}$ ratio is
simply
\begin{equation}
B \, = \, 5 \, U^{1/2}_{\rm ext}
 {\Gamma\over (L_{\rm c}/L_{\rm s})^{1/2}} \, {\rm G}.
\end{equation}
Using Eqs. 3 and 6 we can then constrain the values of $\Gamma$ and
$B$ which minimize the total bulk power and at the same time ensure
the required Compton over synchrotron luminosity ratio.  Note that
$L_{\rm B} \propto \Gamma^4$ while $L_{\rm p,e}\propto \delta^{-4}$.

On the other hand, the Compton scattered radiation field can be
dominated by synchrotron photons. Clearly, this would be favoured (for
a given $B$) at low values of the bulk speed, such that the role of
the external photons (i.e. their comoving energy density) is
diminished in favour of the synchrotron ones (Sikora, Begelman \& Rees
1994). In such situation the Compton dominance is given by
\begin{eqnarray}
{L_{\rm c}\over L_{\rm s}} &=& \, {2\over c R^2} {L_{\rm s}\over
\delta^4 B^2}, 
\end{eqnarray}
which leads to a value for the magnetic field
\begin{equation}
B = 8.2\times 10^{-6} {L^{1/2}_{\rm s}\over 
(L_{\rm c}/L_{\rm s})^{1/2} R \delta^2}
 \, {\rm G}, 
\end{equation}
resulting in $L_{\rm B} \propto \Gamma^2 \delta^{-4}$ and $L_{\rm
p,e}\propto \Gamma^2$.

Note that the dependence of $L_{\rm p,e}$ and $L_{\rm B}$ with respect 
to the bulk Lorentz factor is radically different for these two
possible origins of the high energy component.  
In the following we derive the minimum total power self--consistently, 
by taking into account which radiative process dominates for different 
values of $B$ and $\Gamma$.

\section{Application to PKS 0637--752}

Consider the specific case of PKS 0637--752, for which we have
information on both the sub-pc and the extended jet.

In Celotti et al. (2001) the spectrum of the X--ray knot WK7.8 was
successfully fitted by adopting a one zone, homogeneous synchrotron
and inverse Compton model and assuming relativistic bulk motion.  
As mentioned in the introduction, this was the key to explaining the strong
X--ray emission (because the CMB is seen in the comoving frame boosted
by the factor $\sim\Gamma^2$ and thus dominates over the locally
produced synchrotron energy density, in turn enhancing the resulting
X--ray emission).  
If the knot were not moving relativistically, many more leptons would be 
required to produced the observed X--rays, since in this case: 
i) the flux would not be beamed and the intrinsic emitted luminosity 
would be larger; 
ii) the CMB would not be seen boosted and the inverse Compton process 
on external photons would be less favoured.

Specifically, the model parameters adopted by Celotti et al. (2001)
for the knot emission of this source assumed a bulk Lorentz factor
$\Gamma=14$, a viewing angle $\theta=5^\circ$ and a source dimension
$R=10^{22}$cm.  Note that the emitting region is resolved in the
X--ray images and thus $R$ is not a free parameter. Also the viewing
angle is limited by the VSOP observation of superluminal motion, but
only to be $\theta < 6^\circ$ at the pc scale; as radio maps do not
show any bending of the jet between the pc and $>100$ kpc scales, we
assume the large scale jet to be at a similar viewing angle.

The average $\langle\gamma\rangle$, $\langle\gamma^2\rangle$ derived
by the spectral fitting are reasonably well constrained by the
spectral slope and by interpreting the optical flux as due to
synchrotron (this limits the maximum lepton energy $\gamma_{\rm max}$)
and the X--rays as due to inverse Compton off the CMB.  
This latter constraint limits the minimum Lorentz factor of the
distribution, $\gamma_{\rm min}$, to values smaller than $\sim$30,
thus reducing the (usually large) uncertainty on the total power
associated with this parameter.  
Note that although the particle
distribution is also a function of the assumed $B$ and $\Gamma$, this
dependence has an almost negligible effect on the power estimates
In constructing Fig. 2 and Fig. 3, we
have however taken into account the dependence of
$\langle\gamma^2\rangle$ on $B$ and $\Gamma$ by appropriately scaling
in order to have the correct observed synchrotron peak frequency.

From all the above parameters we estimated the powers discussed in the
previous section for the knot WK7.8 of PKS~0637--752.  The main point
to be stressed is that the total power (dominated by $L_{\rm p}$) has
a minimum for values of the Lorentz factor extremely close to that
determined by the spectral fit ($\Gamma = 11$ vs $\Gamma=14$).

This applies of course for an ordinary e--p plasma as $L_{\rm p}$ is
calculated assuming an equal number of protons and emitting
electrons. At this minimum the contribution of the Poynting flux is
smaller (by a factor $\sim$10) than $L_{\rm p}$ and close to
equipartition with the kinetic power associated with the emitting
particles. Indeed in the case of an e$^{\pm}$ plasma, the total
minimum is reached for $\Gamma \sim 10$ at the equipartition between
$L_{\rm B}$ and $L_{\rm e}$.
 
Given the relative freedom on the viewing angle and the fact that a
decrease in $\theta$ leads to a lowering of the total required power,
we tried to find independent estimates of $L_{\rm tot}$. In particular
we considered the correlation found by Rawlings \& Saunders (1991)
between the luminosity in the narrow line emission and the power
required for jets to supply the extended lobe energy in their
lifetime. From the luminosity in $[OIII]$ (Tadhunter et al. 1993) we
were able to estimate -- following Rawlings \& Saunders' prescription
-- the total luminosity in narrow lines $L_{\rm NLR} \simeq 5 \times
10^{44}$ erg s$^{-1}$.  According to the above correlation this
corresponds to $L_{\rm tot} \sim 10^{47}$ erg s$^{-1}$, although the
correlation has (large!) uncertainties, 
and no energy other than that associated with the 
relativistic emitting particles in the radio lobes is assumed.

\begin{figure}
\psfig{file=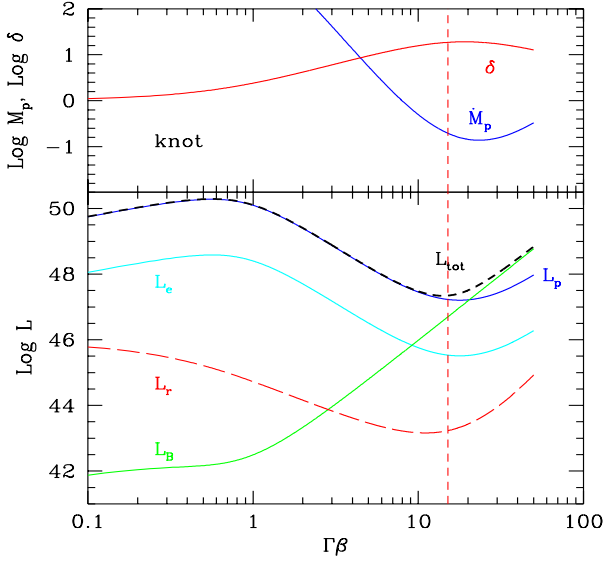,width=9truecm,height=8truecm}
\vskip -0.3 true cm
\caption{Lower panel: The lepton and proton bulk kinetic powers
($L_{\rm e}$ and $L_{\rm p}$ respectively), the Poynting flux ($L_{\rm
B}$), the radiated [$L_{\rm r} \sim (L'_{\rm s} + L'_{\rm c}]
\Gamma^2$) and the total kinetic power ($L_{\rm tot}$) as a function
of $\Gamma$ for the parameters corresponding to the W7.8 knot of PKS
0637--752, as described in the text. The upper panels report the
corresponding mass fluxes $\dot M_{\rm p}$ (in $M_{\odot}$ yr$^{-1}$)
and Doppler factors.  The vertical dashed line corresponds to
$\Gamma=15$, the value used in the spectral fit.  }
\end{figure}

\begin{figure}
\psfig{file=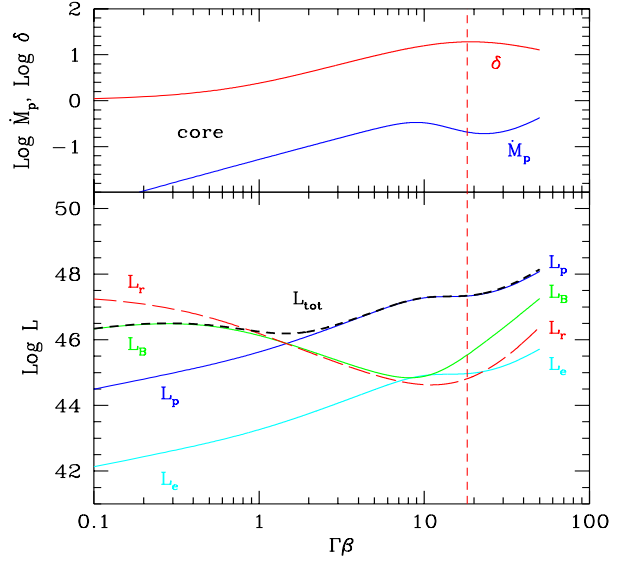,width=9truecm,height=8truecm}
\vskip -0.3 true cm
\caption{As in Fig. 2 for the parameters corresponding to the sub--pc
core of PKS 0637--752, as described in the text.  The vertical dashed
line corresponds to $\Gamma=18$, the value used in the spectral fit.
}
\end{figure}

Consequently we explored the possibility of a viewing angle smaller
than $5^{\circ}$.  
In particular we report here the case of $\theta = 3^{\circ}$, which 
roughly corresponds to a minimum power similar to that determined 
from the lobe energy content and line luminosity. 
The results are graphically displayed in Fig.~2. 
Similarly to the case already discussed the minimum value of the total 
power is reached for a bulk Lorentz factor extremely close to that inferred 
independently from the spectral fit (which required $\Gamma=15$ and 
therefore almost the same $B$--field as before
but a smaller intrinsic luminosity).
In this case $L_{\rm p}$ and $L_{\rm B}$ are closer to equipartition and 
about a factor 8 lower than for $\theta=5^{\circ}$.

We also considered analogous estimates for the core (blazar)
component of PKS~0637--752, assuming $\theta=3^\circ$, $\Gamma=18$, 
$R=2.5\times 10^{16}$ cm, $L_{\rm BLR}=10^{45}$ erg s$^{-1}$, 
$R_{\rm BLR} \sim$ 1 pc, 
$B=2.2$ G, resulting in $U_{\rm ext}\sim 0.1$
and $U_{\rm B}\sim 0.2$ erg cm$^{-3}$.
Results are shown in Fig.~3.  
No consistent solution can be found at low $\Gamma$, where the dissipated
power exceeds the total power carried by the jet.
In addition, $\Gamma$\gtsima 18 is implied by the observed 
superluminal motion.
Close to $\Gamma\sim 18$, 
the total power (once again dominated by $L_{\rm p}$)
presents a flattening (not really a minimum) at a value similar to
that on extended scales and again for a $\Gamma$ close to that
inferred from the spectral fit ($\Gamma=18$), and not far from
equipartition of the magnetic field and lepton components.

The two figures (2 and 3) also show in the upper panels the
corresponding mass fluxes $\dot M_{\rm p}$, estimated assuming one
proton per lepton, and the Doppler factors. Again on both scales a
minimum in $\dot M_{\rm p}$ is reached close to the minimum power
condition and corresponds to values $0.1-0.2 M_{\odot}$ yr$^{-1}$.
This can be compared with the accreting mass flux (note that at small
scales entrainment is not likely to be substantial).  If we consider a
covering factor $\sim 0.01$ of the narrow line emitting gas, the
corresponding ionizing (disc) emission requires an accretion flow of
$\sim 10 M_{\odot}\eta^{-1}_{0.1}$ yr$^{-1}$ for a radiative
efficiency of $\eta = 0.1$: only about 1 per cent of the mass would
need to be channeled and accelerated in jets to ultra--relativistic
speeds.

\section{Discussion}

We have considered the implications of observed spectral quantities,
namely the synchrotron (low energy) and inverse Compton (high energy)
components which dominate the emission 
%
%
of extragalactic jets.  
Besides the total radiated luminosity, these quantities allow to estimate 
the corresponding bulk kinetic and magnetic powers as functions
of the assumed bulk Lorentz factor.

In particular it is possible to perform such estimates
also for large scales jets which has been seen by $Chandra$. 
We have therefore analysed the case of PKS 0637--752, for which
broad band emission has been observed from both the blazar (sub-pc)
and the extended jet.
  
For the range of observing angles consistent with superluminal
constraints we find that: a) the total power is minimized for
relativistic bulk Lorentz factor larger than 10; b) the `fitting' of
the observed spectral energy distribution implies physical parameters
which lie close to this minimum in the total power.  
These
findings somehow lend support to the hypothesis of highly relativistic
motion on the hundred--kiloparsec scale in jets, given the already
high values of the estimated powers, which appear to be close to the
limits of the proposed power production mechanisms
in AGN.  
Note that the possible range of total power spans about
three and the ratio $L_{\rm p}/L_{\rm B}$ about eight orders of
magnitude.  Furthermore even the minimum power inferred (referring
only to the non--thermal emitting jet plasma) is of order 10$^{47}$
erg s$^{-1}$, pushing (once again) the requirement for any powering
mechanism.

Within the acceptable range of viewing angles, a specific choice of
$\theta \sim 3^{\circ}$ provides absolute values for the total power
which well agree with both $L_{\rm tot}$ derived on sub--pc scale and
the jet power estimated from the lobe energy content
(but note that this `solution' implies to double
the already large linear dimension of the jet).
 
In other words it appears that jets are in the most energetically
economic conditions for the given emitted radiation and spectrum, i.e.
that there is a close link between the total power, the
dimension, the particle distribution and the intensity of the
field. Despite of this the inferred radiative efficiency turns out to
be only of the order $\sim 3\times 10^{-3}$ -- $10^{-4}$, for the
nuclear and extended regions, respectively. 
It is worthwhile to point out that such low radiative efficiencies
in the kpc--scale jet, and their higher values on
smaller scales, may be a natural outcome of the internal--shock blazars
scenario (e.g. Ghisellini 2000; Spada et al. 2001).

In these conditions the mass flux in the jet corresponds to about one
per cent of the mass flux accreting on the central source (for
$\eta\sim 0.1$), as deduced from the luminosity of the narrow lines.

It is relevant to comment on the possibility of an e$^{\pm}$ dominated jet.  
Solutions close to the minimum total power for ultra--relativistic 
large (and small) scale jets can also be obtained. 
Furthermore such solutions appear to be intriguingly close
to equipartition values of $L_{\rm e}$ and $L_{\rm B}$. 
However, if the emission line luminosity is a measure of the lobe
energetics, then $L_{\rm e}+L_{\rm B}$ 
is too small to power the radio lobe, implying a contribution
from the proton component, thus limiting the 
maximum pair content to 5--6 pairs per proton.

On large scales the minimum power condition also corresponds to a
situation of quasi--equipartition between the magnetic and kinetic
power transported in the jet. It is therefore interesting to ask
whether there is a simple and robust way for the particle bulk motion
and the magnetic field components to reach it.  Theoretical studies of
relativistic hydromagnetic flows (e.g. Li, Chiueh \& Begelman 1992;
Begelman \& Li 1994 and references therein) have shown that rough
equipartition between particle (kinetic) and Poynting flux is indeed a
quite natural outcome for reasonable assumptions for the boundary
conditions, with Poynting flux gradually accelerating matter.

Clearly, various types of blazars present different spectral
characteristics and radiated luminosities. We are already considering
the implications for the total power involved obtained in the different
cases (Ghisellini \& Celotti 2001, in prep).  A further analysis we
intend to perform is on data relative to the extended emission from
radio galaxies, in particular to estimate the energetics which might
be associated with a slower moving (external) layer.
A final point to be stressed concerning the recent and future $Chandra$
observations is that although the procedure for the determination of 
powers is similar for the small-- and large--scale jets, the 
large--scale powers are much better determined because the energy density 
of the external radiation field is well known.

\section{Acknowledgments}
We thank the anonymous referee for useful comments which helped  
clarifying the paper.
The Italian MURST is acknowledged for financial support (AC).

\section*{References}

\refitem Begelman M.C., Li, Z.-Y., 1994, ApJ, 426, 269

\refitem Bridle A.H., Hough D.H., Lonsdale C.J., Burns J.O. \& Laing R.A., 
1994, AJ, 108, 766

\refitem Burbidge G.R., 1959, ApJ, 129, 849

\refitem Celotti A., Ghisellini G., Chiaberge M., 2001, MNRAS, 321, L1

\refitem Chartas G., et al., 2000, ApJ, 542, 655

\refitem Chiaberge M., Celotti A., Capetti A., Ghisellini G., 2000,
A\&A, 358, 104

\refitem Dermer C. D., 1995, ApJ, 446, L63

\refitem Fanaroff B.L., Riley J. M., 1974, MNRAS, 167, 31

\refitem Ghisellini G., 2000, in Blazars Physics and Demographics,
Urry M.C., Padovani P. eds., Astronomical Society of the Pacific
(ASP), in press (astro--ph/0011356)

\refitem Gliozzi M., Bodo G., Ghisellini G., 1999, MNRAS, 303, L37

\refitem Hardcastle M.J. Alexander P., Pooley G.G. \& Riley J.M., 
1999, MNRAS, 304, 135

\refitem Laing R., 1993, in Astroph. Jets, Burgarella D., Livio M.,
O'Dea C. eds., Cambridge University Press, Cambridge, 95

\refitem Laing R.A., Parma P., de Ruiter H.R., Fanti R., 1999,
MNRAS, 306, 513

\refitem Li Z.-Y., Chiueh T., Begelman M.C., 1992, ApJ, 394, 459

\refitem Lovell .E.J., 2000, in Astroph. Phenomena Revealed by Space
VLBI, Hirabayashi H., et al. eds., (Sagamihara: ISAS), 215

\refitem Rawlings S.G., Saunders R.D.E., 1991, Nature, 349, 138 

\refitem Sikora M., Begelman M.C., Rees, M.J., 1994, ApJ, 421, 153

\refitem Schwartz D.A., et al., 2000, ApJ, 540, L69

\refitem Spada M., Ghisellini G., Lazzati D., Celotti A., 2001, MNRAS,
in press (astro--ph/0103424)

\refitem Tadhunter C.N., Morganti R., di Serego Alighieri S., Fosbury
R.A.E., Danziger I.J., 1993, MNRAS, 263, 999

\refitem Tavecchio F., Maraschi L., Sambruna R.M., Urry C.M., ApJ,
544, L23
 
\refitem Tingay S.J., et al. 1998, ApJ, 497, 594

\refitem Wardle J.F.C. \& Aaron  S.E., 1997, MNRAS, 286, 425

\end{document}